# Multicanonical study of $D = 2$ $O(3)$ nonlinear $\sigma$-model [*]

T. Neuhaus[a],

[a] Fakultät für Physik, Universität Bielefeld, 33615 Bielefeld, Germany.

We present a new and exploratory approach to determine the $\Delta\beta(\beta)$-shift in the $O(3)$ nonlinear $\sigma$-model. The method is based on a scaling hypothesis for a free energy difference, which is assumed to be valid in a situation where the mass gap correlation length $\xi$ is of the order or larger than the linear extent $L$ of the considered square lattice sizes. The free energy difference arises from the finite volume constraint effective potential of the theory. While the constraint effective potential is calculated in numerical simulations employing a variant of the multicanonical ensemble on medium sized lattices, it is possible to estimate $\Delta\beta(\beta)$ up to a value of $\beta = 2.8$ in the standard parameterization of the model.

## 1. INTRODUCTION

The $D = 2$ nonlinear $O(3)$ sigma model with the lattice action

$$S = -\beta \sum_{x,\nu} \Phi_x^\alpha \Phi_{x+\nu}^\alpha \qquad (1)$$

is believed to be a prototype model of asymptotically free theories and therefore deserves special interest in view of asymptotically free gauge theories. It possesses a finite mass gap and in the perturbative regime of couplings $\beta$ its finite mass-gap correlation length $\xi^{-1} = m_{Gap} a$ is supposed to scale like

$$\xi(\beta) = C_\xi \frac{e^{2\pi\beta}}{\beta}(1 + ...). \qquad (2)$$

Possible corrections to this scaling law (...) have been considered in perturbation theory and are small. In this paper we will be concerned with numerical studies of the $\sigma$-model. Over the years there have been several pioneering numerical studies [1] trying to infer the $\Delta\beta(\beta)$-shift of the theory. These calculations located the $\beta$-region, where perturbative scaling is valid, and attempted to determine the mass gap of the theory in physical units. Hereby corresponds $\Delta\beta(\beta)$ to the positive shift in $\beta$ implied by a doubling of the correlation length and its perturbative prediction is

$$\Delta\beta(\beta) = \frac{ln2}{2\pi}(1 + \frac{1}{2\pi\beta} + O(1/\beta^2)). \qquad (3)$$

[*]preprint number : BI-TH 93/72

Most of the numerical simulations have used measurements of two-point correlation functions in order to determine the mass gap [1] and the quality of these simulations was boosted with the invention of the reflection cluster algorithm. Other numerical calculations have resided on the renormalization group approach [2]. Nevertheless the exponential increase of $\xi$ with $\beta$ makes direct and sensible two point function measurements impossible for $\beta$-values above $\beta \approx 2$ on today's computers. With the help of the renormalization group it was possible to extend the accessible $\beta$-region perhaps up to $\beta \approx 2.4$, but still, very large lattices are needed in these simulations.

The constraint effective potential $U_{eff}(M)$ (CEP), of spin models with continuous global $O(N)$ symmetry is a function of the length $M$ of the mean-field

$$M^\alpha = \frac{1}{L^D} \sum_x \Phi^\alpha; \quad M = \sqrt{M^\alpha M^\alpha} \qquad (4)$$

and is defined by rewriting the finite lattice path integral of the theory in the form

$$Z = c \int dM M^{N-1} e^{-U_{eff}(M)}, \qquad (5)$$

which can be achieved in principal by introducing suitable $\delta$-functions. Note that in the integral a term $M^{N-1}$ appears. It counts the degeneracy of states with given $M$ and is proportional to the surface of the $N$-dimensional sphere. The CEP has been extensively studied in another context with analytic methods [3] as well with numerical simulations [4], namely in $D = 3, 4$ $O(N)$-symmetric $\Phi^4$-theories at a value of the bare quartic coupling $\lambda = \infty$. These models are ferromagnets and possess, unlike the $D = 2$ $O(3)$ nonlinear $\sigma$-model, a symmetry broken phase with a nonzero field expectation value $\Sigma$. In the broken phase the CEP has on finite lattices at $M_{min}$ a minimum, $M_{min}$ is finite in the thermodynamic limit, corresponding to $\Sigma$. Here we may note that both analytic as well as the numerical considerations in these model were concerned with the shape of the CEP in the vicinity of its minimum and that not much is known about it in regions of its argument far away from the minimum, e.g. at $M = 0$, see the discussion below.

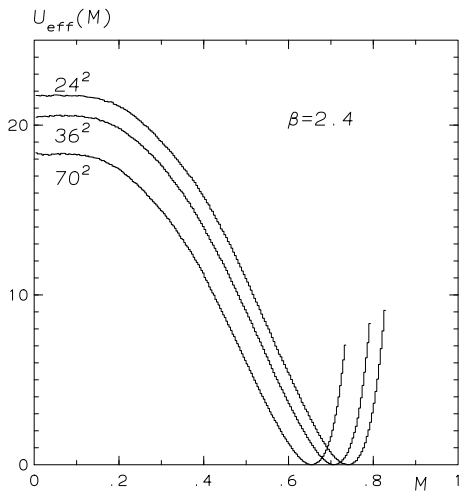

Figure 1. Plot of the constraint effective potential at $\beta = 2.4$ on $24^2, 36^2$ and $70^2$ lattices.

The $D = 2$ $O(3)$ nonlinear $\sigma$-model however exhibits a completely different physical situation. It possesses just a single symmetric phase and therefore an appropiately defined field expectation value $\Sigma$ is identical to zero. Nevertheless, on finite lattices, e.g. square lattices with periodic boundary conditions, one can observe a nonvanishing expectation value of the above operator $< M >$ even for the sigma-model, when the mass gap correlation length $\xi$ is of the order or much larger than the linear extent $L$ of the considered lattices. For purpose of illustration we display here in

Fig. 1 a plot of the CEP at a value of $\beta = 2.4$ on $24^2, 36^2$ and $70^2$ lattices, as it was determined in numerical simulations, which we describe later. We have normalized the value of $U_{eff}(M)$ at its clearly visible minimum to 0. But, how then knows the CEP of the $D = 2$ $\sigma$-model about the symmetric phase of the theory ? Qualitativly we make the following observations: Firstly the location of the minimum $M_{min}$ moves to smaller values if $L$ is increased. Secondly the difference $\Delta U_{eff} = U_{eff}(M=0) - U_{eff}(M_{min})$ decreases when $L$ is increased. This second observation is essential here, as the difference $\Delta U_{eff}$ plays the role of a potential barrier in the CEP for states which, at $M_{min}$ on the finite lattice are probable states, and states, which at $M = 0$ are suppressed in the path integral on finite lattices. These states are however characteristic for the theory in the thermodynamic limit. After all, there we expect a vanishing field expectation value $< M >= 0$. We therefore expect that in this limit

$$\Delta U_{eff} = 0, \quad M_{min} = 0 \qquad (6)$$

for all values of $\beta$ in the $D = 2$ sigma-model. At this point we may come back to above mentioned ferromagnets. There it is expected that

$$\Delta U_{eff} \propto L^{\alpha} \qquad (7)$$

with $\alpha = D - 2$. Thus the volume normalized CEP $U_{eff}/L^D$ of such theories will in the thermodynamic limit have a convex shape with constant values between 0 and some $M_{\Sigma}$.

## 2. SCALING HYPOTHESIS

Considering the path integral of the theory for states with $M^{\alpha} = (M^1, ..., M^N)$ fixed i.e., introducing the constraint partition functions

$$Z(M^{\alpha}) = \int D\Phi \delta^{(N)}(M^{\alpha} - \frac{1}{L^D}\sum_x \Phi^{\alpha})e^{-S} \qquad (8)$$

we note that due to $O(N)$-symmetry these functions are functions of $M$ alone. Furthermore a free energy difference of two such states can be expressed in terms of the CEP in a very simple way

$$\Delta F = -ln\frac{Z(M_1)}{Z(M_2)} = U_{eff}(M_1) - U_{eff}(M_2). \qquad (9)$$

Thus $\Delta U_{eff} = U_{eff}(M=0) - U_{eff}(M_{min})$ corresponds to a free energy difference. In statistical mechanics there exists a powerful scaling hypothesis, Fisher scaling, which is applicable in the framework of finite size scaling theory. At the critical point i.e., in situations where the correlation length $\xi$ is of the order or much larger than any linear extent $L$ of the finite volume system, a free energy or their difference is a function of the ratio $\xi/L$ alone. Similar idears have been presented in [5] for the case of the Ising model and for the helicity modulus of the the $D=2$ $O(3)$ nonlinear $\sigma$ model [6]. In this paper we will test a corresponding scaling hypothesis for the $D=2$ $O(3)$ nonlinear $\sigma$-model i.e., we will assume that

$$\Delta U_{eff} = \Delta U_{eff}(\xi/L) \qquad (10)$$

for values of $\beta$ and lattice sizes $L$ where the control parameter $\xi/L$ is large. Specifically we will assume an analytic form for the $\beta$-dependence of $\xi$ in the vicinity of a given $\beta_0$

$$\xi(\beta) = \xi(\beta_0)\frac{\beta_0}{\beta}e^{\Delta\beta/B_0}; \quad \Delta\beta = \beta - \beta_0 \qquad (11)$$

and we will consider the coefficient $B_0$ as free parameter. $B_0 = 1/(2\pi)$ in perturbation theory. $B_0$ and also the $\Delta\beta(\beta)$-shift are then determined by the behavior of $\Delta U_{eff}$ in the vicinity of $\beta_0$ as determined by the Monte Carlo simulation.

## 3. NUMERICAL SIMULATIONS

The CEP can be obtained in numerical simulations from the probability distribution function $P(M)$, which denotes the probabilty of states with mean field length $M$ in the canonical ensemble

$$e^{-U_{eff}(M)} \propto \frac{P(M)}{M^{N-1}}. \qquad (12)$$

However, simulations of the path integeral on lattice sizes and $\beta$-values considered here strongly suppress states with small values of $M$. We therefore have used a variant of the multicanonical ensemble, in which we have added a weight function $W(M)$, enhancing the appearance of states with small values of $M$ in the simulation. This way it was possible to obtain distribution function estimates for the function $U_{eff}(M)$ on the whole relevant $M$-interval. Typically 400 bins were used for these distribution functions and Fig. 1 shows an example. We have considered four $\beta$-values, namely $\beta = 1.6, 2.0, 2.4$ and $\beta = 2.8$. The considered range of square lattices was between $L = 24$ and $L = 80$. We use periodic boundary conditions. Each simulation was performed with a statistics of about $10^6$ sweeps. Error calculation was performed with the jackknife method. For each pair of values $L$ and $\beta$ we have determined $\Delta U_{eff}$ by fits to the distribution functions. The value of $U_{eff}$ at $M_{min}$ was determined by fitting a parabola to $U_{eff}$ in the vicinity of the minimum, in the same way a polynomial fit was used to determine $U_{eff}(M=0)$.

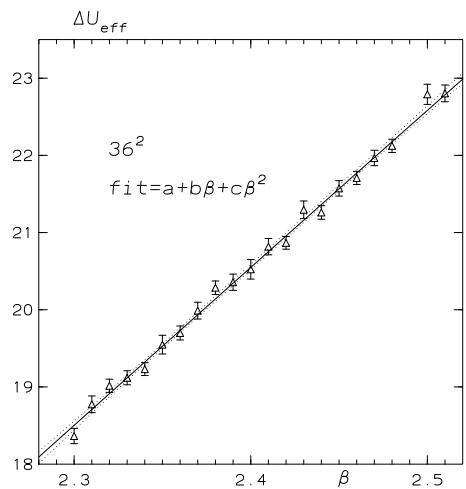

Figure 2. Plot of $\Delta U_{eff}$ for a $36^2$ lattice close to $\beta = 2.4$ as a function of $\beta$. The displayed curve is a polynomial fit to the data.

## 4. RESULTS

Here we refrain from a detailed discussion of our results from simulations at $\beta = 1.6, 2.0$ and $2.8$, the final $\Delta\beta(\beta)$-shift for these $\beta$-values can be found in Fig. 4. Results at these $\beta$-values appear similar to results for $\beta = 2.4$, which we present now. In Fig. 2 we display results for the quantity $\Delta U_{eff}$ as a function of $\beta$ close to $\beta = 2.4$. One observes a monotic increase of $\Delta U_{eff}$ with $\beta$. As can be seen, the functional dependence on $\beta$ is very smooth and therefore a polynomial approximation of the form

$$\Delta U_{eff} = a + b(\beta - \beta_0) + c(\beta - \beta_0)^2; \quad \beta_0 = 2.4 \quad (13)$$

gives a very good description of the analytic behavior with $\beta$. This analytic form can then be employed in order to test the scaling hypothesis Eqs. 10, 11. Figure 3 displays the $L$-dependence of $\Delta U_{eff}$ at $\beta = 2.4$. Lattice sizes range inbetween $L = 24$ and $L = 70$. The decrease of $\Delta U_{eff}$ with increasing $L$ is observed. The curve in Fig. 3 corresponds to the rescaled data on $36^2$ lattices, Eq. 13 and Fig. 2. We have used the scaling hypothesis $\Delta U_{eff} = \Delta U_{eff}(\xi/L)$ and the $\xi(\beta)$ dependence Eq. 11, with a coefficient $B_0$ as determined by a $\chi^2$-fit to the data. We note that the scaling hypothesis is well satisfied. Figure 4 then contains the inferred $\Delta\beta(\beta)$-shifts for all our $\beta$-values.

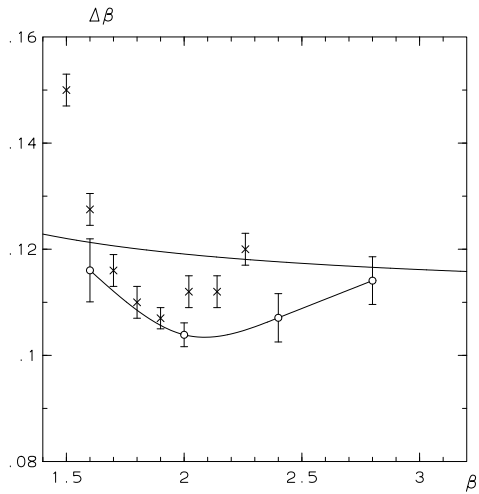

Figure 4. The $\Delta\beta(\beta)$-shift (circles) as determined by our analysis. The curve corresponds to the perturbative result Eq. 3. Circles have been connected with a spline. Crosses are from [2].

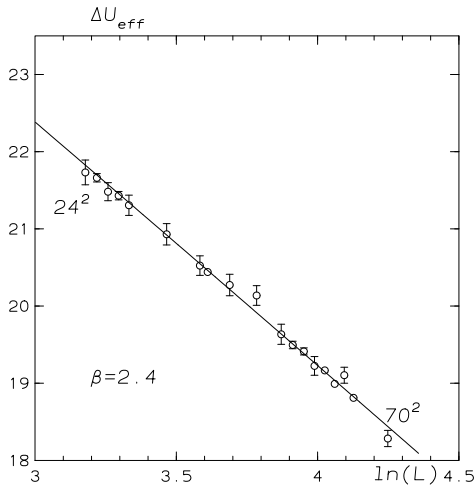

Figure 3. Plot of $\Delta U_{eff}$ as a function of ln(L) at $\beta = 2.4$. The curve corresponds to the rescaled data of Fig. 2 using the scaling hypothesis Eq. 10 and Eq. 11.

## 5. DISCUSSION

Based on a scaling hypothesis for a free energy difference we have estimated the $\Delta\beta(\beta)$-shift in the $D = 2$ $O(3)$-nonlinear $\sigma$-model, see Fig. 4, for $\beta$-values as large as $\beta = 2.8$ on medium sized lattices. While at small values of $\beta$ we find agreement with other numerical determinations, our calculation indicates a slower approach to perturbative scaling, than previously acknowledged. It would be very helpful, if one could control scaling relations like Eqs. 10, 11 and possible scaling deviations from them, with the help of analytic calculations for $U_{eff}(M)$ in the future. If these calculations would confirm our assumptions, then possible benefits are significant, as one might then more closely approach the perturbative limit of asymptotically free theories in numerical simulations.